\documentclass{Interspeech}

\interspeechcameraready

\usepackage[activate]{microtype}
\title{Quantifying Dimensional Independence in Speech: An 
Information-Theoretic Framework for Disentangled Representation Learning
}

\author[affiliation={1}]{Bipasha}{Kashyap}
\author[affiliation={2,3}]{Björn W.}{Schuller}
\author[affiliation={1}]{Pubudu N.}{Pathirana}

\affiliation{NSBE Research Lab, School of Engineering}
            {Deakin University}
            {Australia}

\affiliation{Chair of Health Informatics (CHI)}
            {TUM University Hospital}
            {Germany}

\affiliation{Group on Language, Audio \& Music (GLAM)}
            {Imperial College London}
            {UK}

\email{b.kashyap@deakin.edu.au, schuller@tum.de, pubudu@deakin.edu.au}

\usepackage{comment}
\usepackage{graphicx}
\usepackage{tikz}

\begin{document}

\maketitle
\renewcommand\thefootnote{}
\footnotetext{Submitted to INTERSPEECH 2026.}
\renewcommand\thefootnote{\arabic{footnote}}

\begin{abstract}

Speech signals encode emotional, linguistic, and pathological information within a shared acoustic channel; however, disentanglement is typically assessed indirectly through downstream task performance. We introduce an information-theoretic framework to quantify cross-dimension statistical dependence in handcrafted acoustic features by integrating bounded neural mutual information (MI) estimation with non-parametric validation. Across six corpora, cross-dimension MI remains low, with tight estimation bounds ($< 0.15$ nats), indicating weak statistical coupling in the data considered, whereas Source–Filter MI is substantially higher (0.47 nats). Attribution analysis, defined as the proportion of total MI attributable to source versus filter components, reveals source dominance for emotional dimensions (80\%) and filter dominance for linguistic and pathological dimensions (60\% and 58\%, respectively). These findings provide a principled framework for quantifying dimensional independence in speech.
\end{abstract}

\section{Introduction}

Human speech signals simultaneously convey emotional state, linguistic content, and potential pathological markers through a shared acoustic channel. Disentangling these dimensions is essential for downstream applications such as emotion recognition \cite{cho2025diemo}, speaker verification \cite{li2023mutual}, and clinical speech assessment \cite{chang2023spin}. However, existing approaches lack principled measures of disentanglement quality and instead rely on downstream task performance as a proxy indicator of separation.

The fundamental question remains: \textit{to what degree are emotional, linguistic, and pathological speech dimensions statistically independent?} If these dimensions exhibit high mutual information, complete disentanglement may be theoretically infeasible. Conversely, low mutual information suggests that appropriate inductive biases and representation constraints could enable effective separation. Importantly, such independence properties must hold across linguistic and cultural variation if disentangled representations are to generalise across diverse speaker populations, including native and non-native speakers, multilingual communities, and individuals with speech disorders.

We address this question using an information-theoretic framework integrating neural mutual information (MI) estimation with classical source–filter decomposition. Unlike correlation-based metrics, mutual information captures non-linear dependencies, providing a principled basis for disentanglement evaluation \cite{belghazi2018mine}.


Prior work in \textit{speech disentanglement} has focused primarily on pairwise separation problems, most commonly speaker and emotion. Cho et al.\ 
\cite{cho2025diemo} proposed a self-supervised distillation framework for cross-speaker emotion transfer, while Li et al.\ \cite{li2023mutual} employed mutual information minimisation to learn domain-invariant speaker embeddings. Chang et al.\ \cite{chang2023spin} introduced speaker-invariant clustering for content representation learning. While these studies demonstrate progress in isolating specific factors, they do not address the full three-dimensional structure formed by emotional, linguistic, and pathological information streams.

Neural \textit{mutual information estimation} has advanced substantially in recent years. Mutual Information Neural Estimation (MINE) \cite{belghazi2018mine} provides a variational lower bound, while Contrastive Log-ratio Upper Bound (CLUB) \cite{cheng2020club} offers a contrastive upper bound. Jointly applying these estimators enables tighter bracketing of true mutual information than either method alone. In this work, we extend bounded MI estimation with a non-parametric validation estimator, Kraskov-Stögbauer-Grassberger (KSG) \cite{kraskov2004estimating}, to the analysis of multidimensional speech structure, enabling principled quantification of independence across speech information domains.

\noindent Our contributions are summarised as follows:

\begin{enumerate}
\item A bounded mutual information estimation framework that combines neural estimators, MINE and CLUB, with non-parametric validation, KSG, incorporating exponential moving average (EMA) stabilised training and adaptive uncertainty weighting;
\item Empirical analysis showing low cross-dimension mutual information ($<0.15$ nats) in the considered datasets, consistent with weak statistical dependence between feature sets.
\item A source and filter attribution analysis revealing dimension-specific acoustic substrates.
\item Cross-corpus validation with estimation uncertainty $<$0.35 nats.

\end{enumerate}
\vspace{-4mm}

\section{Methodology}
\subsection{Problem Formulation}
\begin{figure}[t]
\centering

\begin{tikzpicture}
        \node[draw=gray, rounded corners=5pt, inner sep=1.5pt] {
            \includegraphics[width=0.9\linewidth]{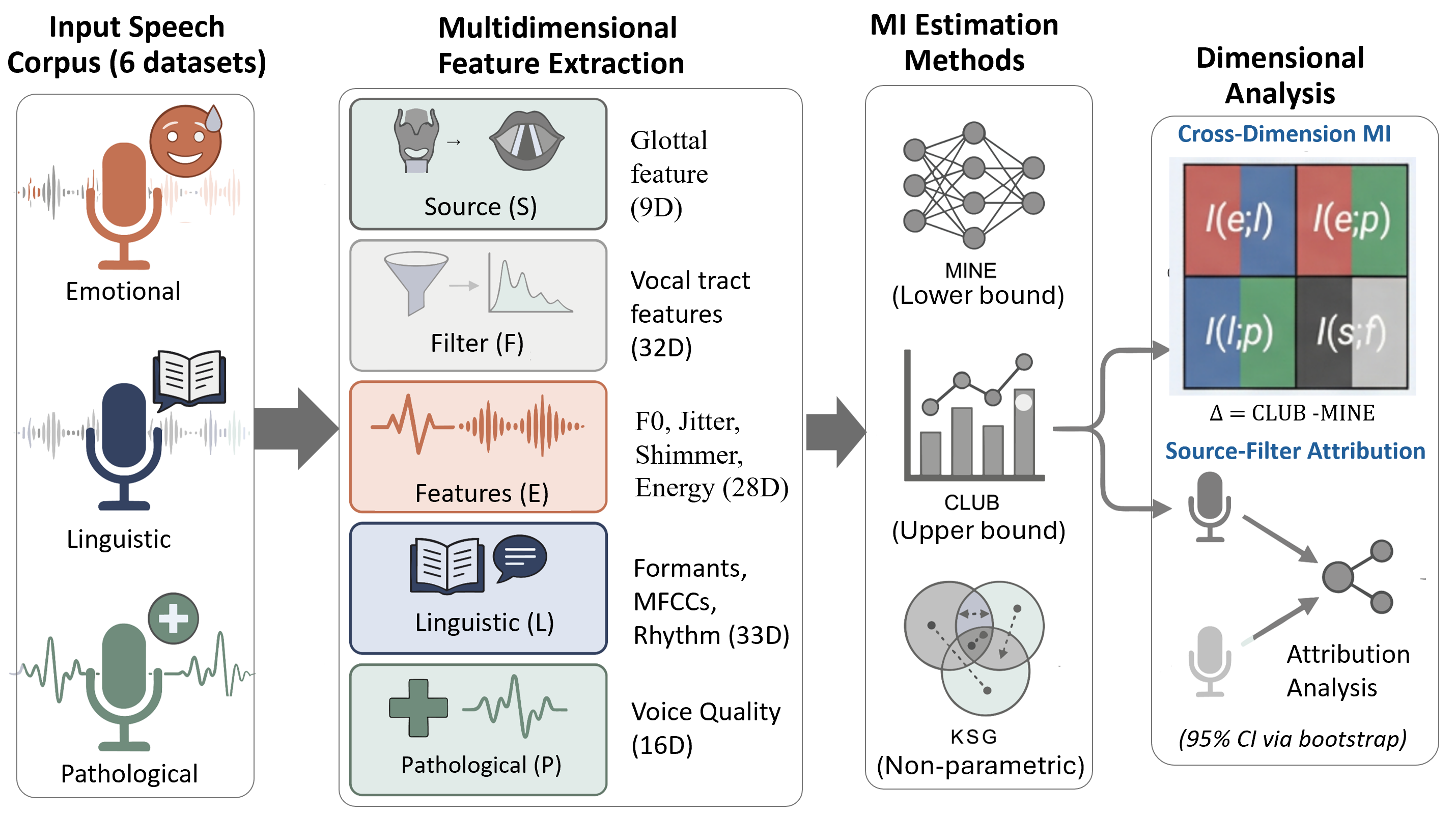}
        };
\end{tikzpicture}
\caption{Proposed framework for quantifying dimensional independence.}
\label{figure 1}
\vspace{-3em}
\end{figure}

\vspace{-2mm}
Let the 
speech signal $\mathbf{x}$ be characterised by semantic dimensions: emotional $\mathbf{e} \in \mathbb{R}^{28}$, linguistic $\mathbf{l} \in \mathbb{R}^{33}$, and pathological $\mathbf{p} \in \mathbb{R}^{16}$. 
The dimension numbers reflect the number of handcrafted acoustic descriptors assigned to each group (see Section~3.2 for the complete feature breakdown). Following the source-filter model \cite{fant1970acoustic}, we extract source $\mathbf{s} \in \mathbb{R}^{9}$ (glottal) and filter $\mathbf{f} \in \mathbb{R}^{32}$ (vocal tract) features. We aim to quantify pairwise MI: $I(\mathbf{e};\mathbf{l})$, $I(\mathbf{e};\mathbf{p})$, $I(\mathbf{l};\mathbf{p})$, and $I(\mathbf{s};\mathbf{f})$. The proposed framework is shown in Figure \ref{figure 1}.

\vspace{-2mm}
\subsection{Feature Extraction}
\vspace{-2mm}
For every audio file, we extract five feature sets using Praat \cite{boersma2001praat}, librosa \cite{mcfee2015librosa}, and openSMILE \cite{eyben2010opensmile} for robust acoustic descriptor computation. 

Acoustic features are inherently multi-functional, such as jitter and shimmer serving as affective and clinical markers, and formants encoding both phonetic identity and articulatory health \cite{schuller2018speech,kent2000research}. Therefore, the semantic groupings below represent \textit{operationally defined feature sets} rather than strictly separable acoustic domains. The MI analysis quantifies the dependency between these defined feature sets.

\textbf{Source features set} $\mathbf{s} \in \mathbb{R}^{9}$ capture glottal characteristics: F0 statistics (mean, std, range, median, Q1, Q3) and voice quality measures (jitter, shimmer, HNR) reflecting laryngeal function \cite{teixeira2013jitter}. \textbf{Filter features set} $\mathbf{f} \in \mathbb{R}^{32}$ capture vocal tract resonances: formant frequencies (F1, F2, F3) and bandwidths (B1, B2, B3), via Burg LPC analysis, and 13 MFCCs with delta coefficients \cite{davis1980comparison}. \textbf{Emotional features} $\mathbf{e} \in \mathbb{R}^{28}$ combine source features with energy contours (RMS mean, std, max) and spectral descriptors (centroid, flux, rolloff). \textbf{Linguistic features set} $\mathbf{l} \in \mathbb{R}^{33}$ combine filter features with delta-delta MFCCs and temporal parameters (speaking rate, duration), capturing primarily phonetic content 
\cite{huang2001spoken}. \noindent\textbf{Pathological features set} $\mathbf{p} \in \mathbb{R}^{16}$ target 
mostly 
%
clinical markers: voice quality measures, formant stability, and F2 transition velocity, a key indicator of articulatory precision in motor speech disorders \cite{skodda2011speech}. The partial feature overlap with source and filter sets (e.g., shared formant and voice quality descriptors) is by design, as it enables the MI framework to quantify precisely how much shared information these operationally distinct groupings carry.

 


\subsection{Bounded Neural MI Estimation}
\vspace{-2mm}
Direct MI computation is intractable for high-dimensional continuous distributions. We employ three complementary estimators providing bounds from multiple directions.

\vspace{-2mm}
\subsubsection{MINE with EMA Stabilisation}
\vspace{-2mm}
The Mutual Information Neural Estimator (MINE) 
\cite{belghazi2018mine} provides a lower bound via the Donsker-Varadhan representation. For any pair of feature vectors $\mathbf{X} \in \mathbb{R}^{d_x}$ and $\mathbf{Y} \in \mathbb{R}^{d_y}$:
\vspace{-2mm}
\begin{equation}
I(\mathbf{X};\mathbf{Y}) \geq \mathbb{E}_{p(\mathbf{x},\mathbf{y})}[T_\theta(\mathbf{x},\mathbf{y})] - \log\mathbb{E}_{p(\mathbf{x})p(\mathbf{y})}[e^{T_\theta(\mathbf{x},\mathbf{y})}],
\label{eq:mine}
\end{equation}
where $T_\theta: \mathbb{R}^{d_x} \times \mathbb{R}^{d_y} \rightarrow \mathbb{R}$ is a critic network. We parameterise $T_\theta$ as a 2-layer MLP with LayerNorm and LeakyReLU:
\vspace{-2mm}
\begin{equation}
T_\theta(\mathbf{x},\mathbf{y}) = \mathbf{W}_2 \cdot \text{LN}(\text{LReLU}(\mathbf{W}_1[\mathbf{x};\mathbf{y}] + \mathbf{b}_1)) + \mathbf{b}_2,
\end{equation}
with hidden dimension $h=256$ and LeakyReLU slope 0.2.

To address high-variance gradients in the partition function estimate, we employ EMA stabilisation over training epochs $t$:
\vspace{-2mm}
\begin{equation}
\bar{Z}_t = (1-\alpha)\bar{Z}_{t-1} + \alpha \cdot \mathbb{E}_{\text{batch}}[e^{T_\theta}],
\end{equation}
with $\alpha=0.01$ and bias correction $\hat{Z}_t = \bar{Z}_t / (1-(1-\alpha)^t)$. The stabilised MINE objective becomes:
\vspace{-3mm}
\begin{equation}
\hat{I}_{\text{MINE}} = \mathbb{E}_{p(\mathbf{x},\mathbf{y})}[T_\theta] - \log(\hat{Z}_t + \epsilon),
\end{equation}
where $\epsilon=10^{-8}$ prevents numerical underflow.

\vspace{-2mm}
\subsubsection{CLUB with Variance Clamping}
\vspace{-2mm}
The Contrastive Log-ratio Upper Bound (CLUB)\cite{cheng2020club} provides:
\vspace{-3mm}
\begin{equation}
I(\mathbf{X};\mathbf{Y}) \leq \mathbb{E}_{p(\mathbf{x},\mathbf{y})}[\log q_\phi(\mathbf{y}|\mathbf{x})] - \mathbb{E}_{p(\mathbf{x})p(\mathbf{y})}[\log q_\phi(\mathbf{y}|\mathbf{x})],
\label{eq:club}
\end{equation}
with equality when $q_\phi(\mathbf{y}|\mathbf{x}) = p(\mathbf{y}|\mathbf{x})$. We model $q_\phi$ as a diagonal Gaussian:
\vspace{-4mm}
\begin{equation}
q_\phi(\mathbf{y}|\mathbf{x}) = \mathcal{N}(\mathbf{y}; \boldsymbol{\mu}_\phi(\mathbf{x}), \text{diag}(\boldsymbol{\sigma}^2_\phi(\mathbf{x}))),
\end{equation}
where $\boldsymbol{\mu}_\phi$ and $\log\boldsymbol{\sigma}^2_\phi$ are MLP outputs. The log-likelihood decomposes as:
\vspace{-3.5mm}
\begin{equation}
\log q_\phi(\mathbf{y}|\mathbf{x}) = -\frac{1}{2}\sum_{j=1}^{d_y}\left[\log\sigma^2_{\phi,j} + \frac{(y_j - \mu_{\phi,j})^2}{\sigma^2_{\phi,j}}\right].
\end{equation}
\vspace{-0.5mm}
To prevent variance collapse or explosion, we clamp log-variance:
\vspace{-3mm}
\begin{equation}
\log\hat{\sigma}^2_\phi = \text{clamp}(\log\sigma^2_\phi, -6, 2),
\end{equation}
constraining variance to $[e^{-6}, e^2] \approx [0.002, 7.4]$.

\vspace{-2mm}
\subsubsection{KSG Non-Parametric Validation}
\vspace{-2mm}
The Kraskov-Stögbauer-Grassberger (KSG)
estimator \cite{kraskov2004estimating} provides training-free validation using $k$-nearest neighbour statistics:
\vspace{-2mm}
\begin{equation}
\hat{I}_{\text{KSG}} = \psi(k) + \psi(N) - \langle\psi(n_x(i) + 1) + \psi(n_y(i) + 1)\rangle_i,
\label{eq:ksg}
\end{equation}
where $\psi(\cdot)$ is the digamma function, $k=5$ is the neighbour count, $N$ is the number of samples, and $n_x(i), n_y(i)$ are marginal neighbour counts within the $k$-th neighbour distance $\epsilon_i$ computed using the Chebyshev ($L^\infty$) norm. We use kd-trees for efficient nearest-neighbour search with $O(N \log N)$ tree construction.

\vspace{-2mm}
\subsubsection{Combined Estimation with Adaptive Weighting}
\vspace{-2mm}
We enforce theoretical bounds $\hat{I}_{\text{MINE}} \leq \hat{I}_{\text{CLUB}}$ and compute uncertainty:
\vspace{-3mm}
\begin{equation}
\Delta = \hat{I}_{\text{CLUB}} - \hat{I}_{\text{MINE}}.
\end{equation}

The final MI estimate uses adaptive KSG-anchored weighting:
\vspace{-3mm}
\begin{equation}
\hat{I}_{\text{final}} = (1-w)\cdot\frac{\hat{I}_{\text{MINE}} + \hat{I}_{\text{CLUB}}}{2} + w \cdot \hat{I}_{\text{KSG}},
\label{eq:final}
\end{equation}
where the weight $w$ increases with uncertainty:
\begin{equation}
w = \begin{cases}
0.3 & \text{if } \Delta \leq 1.0 \\
\min(0.6, 0.3 + 0.1\Delta) & \text{if } \Delta > 1.0.
\end{cases}
\end{equation}
This favours neural estimates when bounds are tight, but anchors to KSG when uncertainty is high.

\vspace{-2mm}
\subsection{Training Protocol}
\vspace{-2mm}
We employ ensemble training with $M=3$ independent estimator pairs per configuration. Each estimator uses Adam optimisation ($\eta=10^{-4}$, weight decay $10^{-5}$) with gradient clipping (max norm 1.0) and ReduceLROnPlateau scheduling (factor 0.5, patience 10). Early stopping triggers when $\Delta < 0.1$ for 7+ consecutive epochs. Final estimates average the last 10 of $T$ training epochs ($T \leq 100$) across ensemble members, where $\hat{I}^{(m)}_t$ denotes the MI estimate from ensemble member $m$ at epoch $t$:
\vspace{-3mm}
\begin{equation}
\hat{I} = \frac{1}{M}\sum_{m=1}^{M}\frac{1}{10}\sum_{t=T-9}^{T}\hat{I}^{(m)}_t.
\end{equation}

\vspace{-2mm}
\subsection{Source-Filter Attribution}
\vspace{-2mm}
For each semantic dimension $\mathbf{d} \in \{\mathbf{e}, \mathbf{l}, \mathbf{p}\}$, we decompose MI into source and filter contributions using KSG estimates. The attribution ratio quantifies the proportion of total MI (source + filter) carried by each production component:
\vspace{-3mm}
\begin{equation}
A_{\text{source}}(\mathbf{d}) = \frac{\hat{I}_{\text{KSG}}(\mathbf{s};\mathbf{d})}{\hat{I}_{\text{KSG}}(\mathbf{s};\mathbf{d}) + \hat{I}_{\text{KSG}}(\mathbf{f};\mathbf{d})},
\end{equation}
with $A_{\text{filter}}(\mathbf{d}) = 1 - A_{\text{source}}(\mathbf{d})$. Unlike variance decomposition or Shapley attribution, this ratio reflects the relative information shared between each production subsystem and a given semantic dimension. Bootstrap resampling ($B=10$) provides confidence intervals.



\vspace{-3mm}
\section{Experimental Setup}
\vspace{-2mm}
\subsection{Datasets}
\vspace{-1mm}
We evaluate the proposed framework across six English speech corpora spanning three dimensions. \textit{Emotional}: RAVDESS \cite{livingstone2018ravdess} (1,440 utterances, 24 actors, eight emotions) and IEMOCAP \cite{busso2008iemocap} (10,039 utterances, scripted and improvised interactions). \textit{Linguistic}: L2-ARCTIC \cite{zhao2018l2arctic} (26,867 utterances, 24 non-native speakers, six native language backgrounds) and the GMU Speech Accent Archive (2,140 speakers representing 177 native languages, all producing a standard English paragraph). \textit{Pathological}: UA-Speech \cite{kim2008uaspeech} (dysarthric English speech, 15 speakers) and MDVR-KCL (Parkinson’s disease English speech via mobile recordings). To ensure robustness and generalisability within the available data, we generate all eight dataset combinations (2$\times$2$\times$2), validating consistency across corpus pairings beyond any single accent, affective, or clinical cohort.
\vspace{-2mm}



\subsection{Implementation Details}
\vspace{-1mm}
Both the MINE and CLUB networks use 2-layer MLPs with 256 hidden units and ReLU activations. Training uses the 
Adam optimiser with learning rate $10^{-4}$ for 100 epochs. Features are z-score normalised. We use 500 samples per dataset,
stratified to ensure balanced representation across speakers and conditions within each corpus, preventing degenerate sampling from single speakers or language backgrounds.
\vspace{-2mm}

\begin{table}[h]
\centering
\caption{Cross-dimension mean MI analysis. MINE = lower bound, CLUB = upper bound, KSG = non-parametric $k$-NN estimator, Final = KSG-anchored weighted estimate (Eq.~\ref{eq:final}). All values in nats. $\Delta$ = CLUB $-$ MINE measures estimation uncertainty. MI $<$ 0.15 nats indicates near-independence.}
\label{table 2}
\vspace{-2mm}
\resizebox{1\columnwidth}{!}{
\begin{tabular}{lcccccc}
\toprule
\textbf{Pair} 
& \textbf{MINE} 
& \textbf{CLUB} 
& $\boldsymbol{\Delta}$ 
& \textbf{KSG} 
& \textbf{Final (Mean $\pm$ Std)} \\
\midrule
Emotion--Linguistic   & 0.00 & 0.14 & 0.14 & 0.25 & \textbf{0.12 $\pm$ 0.05} \\
Emotion--Pathology    & 0.00 & 0.07 & 0.07 & 0.26 & \textbf{0.10 $\pm$ 0.06} \\
Linguistic--Pathology & 0.00 & 0.10 & 0.10 & 0.21 & \textbf{0.10 $\pm$ 0.05} \\
Source--Filter        & 0.24 & 0.59 & 0.35 & 0.60 & \textbf{0.47 $\pm$ 0.38} \\
\bottomrule
\end{tabular}}
\vspace{-3mm}
\end{table}
\vspace{-3mm}

\section{Results}

\subsection{Cross-Dimension Mutual Information}
\vspace{-2mm}
Table~\ref{table 2} aggregates across all dataset combinations: it reports mean MI for each pair across all estimators (MINE, CLUB, KSG, Final), plus mean uncertainty. All cross-dimension pairs exhibit mean MI (Final) below 0.15 nats with tight uncertainty bounds ($\Delta < 0.15$ nats), indicating near-independence in the examined feature space.

\noindent\textbf{Key observations: }\textit{(i) Near-Zero Cross-Dimension MI.} The Emotion--Pathology and Linguistic--Pathology pairs exhibit the lowest MI (0.10 nats), suggesting these dimensions occupy nearly independent subspaces in acoustic feature space. Emotion--Linguistic shows slightly higher coupling (0.12 nats), likely reflecting shared prosodic components (e.g., F0 contours influencing both emotional expression and linguistic stress patterns \cite{schuller2018speech}). However, this coupling remains negligible.
\textit{(ii) Bounded Estimation Validity.} All estimates satisfy MINE $\leq$ CLUB, confirming theoretical validity. The tight bounds ($\Delta < 0.15$ for cross-dimension pairs) indicate that both estimators converged to similar values, providing high confidence in the estimates. KSG values closely match the neural combined estimates, providing independent validation. \textit{(iii) Source-Filter Coupling.} Source-Filter MI (Final, 0.47 nats) is substantially higher than cross-dimension pairs but lower than classical acoustic theory might predict. This suggests that even production-level features maintain quasi-independence when properly extracted, supporting the source-filter decomposition assumption.

\begin{figure}[t]
\centering
\begin{tikzpicture}
        \node[draw=gray, rounded corners=5pt, inner sep=1.5pt] {
            \includegraphics[width=0.9\linewidth]{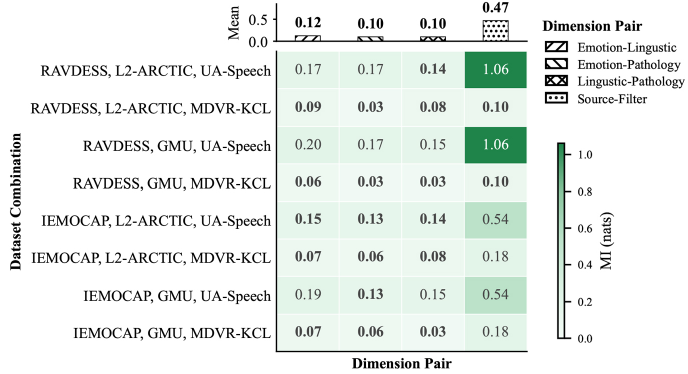}
        };
\end{tikzpicture}
\vspace{-2mm}
\caption{Cross-dimension MI (Final) heatmap across dataset combinations. Lower values (lighter) indicate greater independence between feature sets. Top marginal bars show per-pair means averaged across all combinations.}
\label{figure 2}
\vspace{-2em}
\end{figure}

\vspace{-2mm}
\begin{figure}[h!]
\centering
\begin{tikzpicture}
        \node[draw=gray, rounded corners=5pt, inner sep=1.5pt] {
            \includegraphics[width=0.95\linewidth]{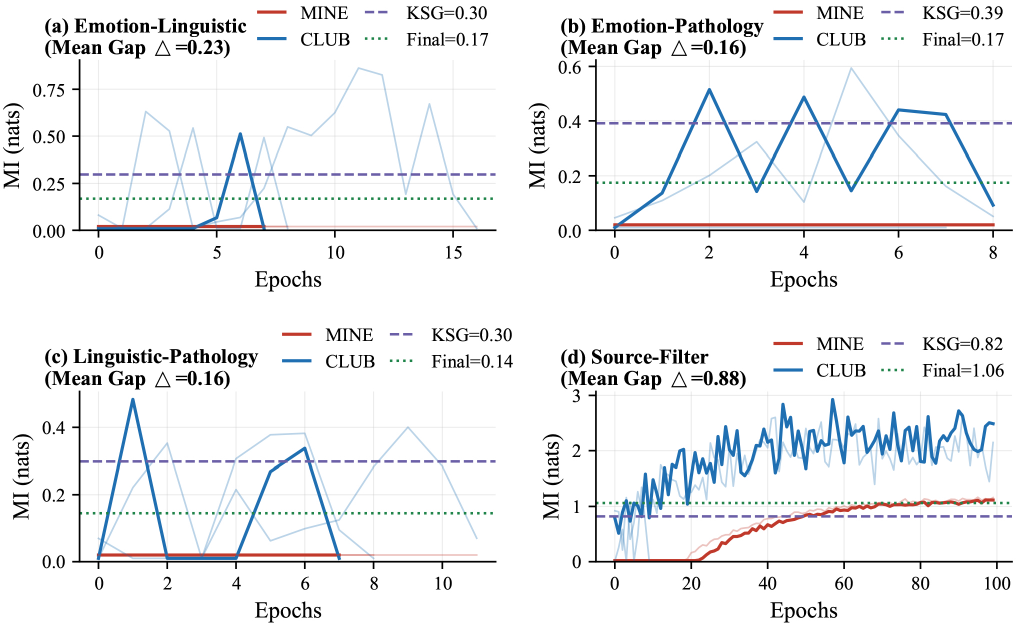}
        };
\end{tikzpicture}
\vspace{-3mm}
\caption{MI estimation convergence across dimensions: (a) Emotion–Linguistic, (b) Emotion–Pathology, (c) Linguistic–Pathology, and (d) Source–Filter. Per-ensemble MINE and CLUB trajectories are shown (offset for visual clarity), alongside the KSG baseline (dashed) and the final consensus estimate for a representative dataset combination (RAVDESS, L2-ARCTIC, UA-Speech). The mean estimator gap ($\Delta$) over the final 10 training epochs is reported, demonstrating progressive reduction and convergence over training.
}
\label{figure 3}
\vspace{-2em}
\end{figure}

\subsection{Cross-Corpus Consistency}
\vspace{-2mm}
Figure~\ref{figure 2} shows MI (Final) across all 8 dataset combinations. Cross-dimension pairs consistently show very low MI (0.03--0.20 nats) while Source-Filter shows higher but stable coupling (0.1--1.06 nats). The consistency across combinations suggests that the observed independence patterns are not artefacts of specific corpus pairings. Standard deviations remain below 0.1 nats for cross-dimension estimates, whereas Source-Filter shows a higher value (0.38) (Table~\ref{table 2}) reflecting dataset-dependent coupling differences, likely driven by recording conditions and speaker populations.

\vspace{-2mm}
\subsection{Convergence Analysis}
\vspace{-2.5mm}
Figure~\ref{figure 3} reveals distinct convergence patterns for one dataset combination (RAVDESS, L2-
ARCTIC, UA-Speech). Cross-dimension pairs converge rapidly (8--16 epochs) with early stopping triggered by $\Delta < 0.1$, consistent with low true MI making the estimation problem easier. Source-Filter requires full training (100 epochs) with continued refinement, reflecting higher underlying dependence.

\vspace{-2.5mm}
\subsection{Source-Filter Attribution}
\begin{figure}[h]
\centering
\begin{tikzpicture}
        \node[draw=gray, rounded corners=5pt, inner sep=1.5pt] {
            \includegraphics[width=0.7\linewidth]{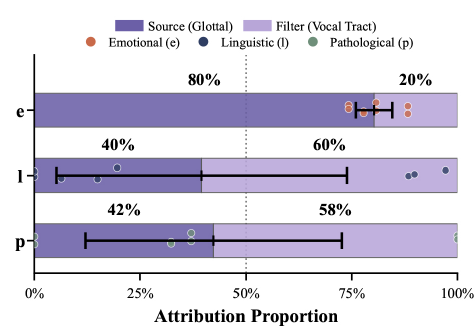}
        };
\end{tikzpicture}
\vspace{-2mm}
\caption{Source–Filter attribution across semantic dimensions. Stacked bars show the mean proportion of MI carried by Source vs.\ Filter components (Eq.~3); individual dataset combinations are overlaid as jitter points. Error bars indicate 95\% CI.}
\label{figure 4}
\vspace{-2em}
\end{figure}

\vspace{-2.5mm}
Figure~\ref{figure 4} reveals dimension-specific acoustic substrates. Values denote the proportion of mutual information attributed to source versus filter components. \textit{Emotional features} are consistently source-dominated across all eight dataset combinations (80\%, 95\% CI [0.76, 0.85]), consistent with the role of F0, jitter, shimmer, and voice quality in affect encoding \cite{scherer2003vocal}. In contrast, \textit{linguistic features} show filter dominance (60\%), reflecting the articulatory basis of phonetic contrasts via formant structure and spectral shaping. \textit{Pathological features} similarly exhibit filter dominance (58\%), suggesting articulatory impairment contributes slightly more than laryngeal dysfunction, consistent with clinical observations of dysarthria \cite{kent2000research,rusz2011acoustic,kashyap2020prominence,kashyap2020cepstrum}. While emotional dimensions remain stably source-driven, linguistic and pathological dimensions display greater inter-combination variability.
\vspace{-3.5mm}

\vspace{-1mm}

\section{Discussion}
\subsection{Theoretical Implications}
\vspace{-2mm}
Our bounded mutual information estimation framework reveals consistently low cross-dimension MI (below 0.15 nats) in handcrafted acoustic feature space, substantially lower than the 0.69--1.96 nats reported in prior work using conventional or single-estimator approaches that can be biased or variance-unstable in high-dimensional settings \cite{belghazi2018mine,cheng2020club}. These near-zero residual dependencies carry important implications.

\noindent\textbf{Disentanglement Feasibility.} Low MI upper bounds (CLUB $<$ 0.15 nats) indicate that near-perfect dimensional disentanglement is theoretically achievable for these feature sets, with negligible irreducible shared information even under a worst-case interpretation.
\textbf{Information-Theoretic Interpretation.} An MI of $\sim$0.10 nats implies that knowledge of one dimension (e.g., pathology) provides negligible predictive information about another dimension (e.g., linguistic or emotion), supporting assumptions of near-independent modelling in practical speech systems.

\vspace{-2mm}
\subsection{Methodological Contributions}
\vspace{-2mm}
The proposed framework addresses known limitations of neural MI estimators.

\noindent\textbf{EMA Stabilisation.} Bias-corrected exponential moving averages stabilise the MINE partition function, reducing gradient variance while preserving asymptotic unbiasedness.
\textbf{Variance Clamping.} Constraining CLUB log-variance to $[-6, 2]$ prevents divergence in low-MI regimes where conditional and marginal distributions converge.
\textbf{Adaptive Anchoring.} KSG-anchored weighting improves robustness when neural bounds diverge, prioritising non-parametric estimates under high estimator disagreement.

\vspace{-2mm}
\subsection{Implications for Encoder Design}
\vspace{-2mm}
The low cross-dimension MI observed in handcrafted features motivates several design principles for speech encoders, pending validation on learned representations.

\noindent\textbf{Separate Pathways.} Near-zero cross-dimension MI ($<$0.1 nats) supports dimension-specific encoder branches with minimal information sharing.
\textbf{Feature Alignment.} Attribution analysis suggests linguistic encoders should prioritise filter features (formants, spectral structure), while emotional encoders benefit from source-dominated cues (F0, voice quality).
\textbf{Regularisation Targets.} Empirical MI baselines ($\sim$0.10 nats) provide realistic disentanglement regularisation objectives.
\textbf{Cross-Lingual Robustness.} Consistency across accent-diversity corpora covering over 177 language backgrounds provides initial evidence for language-robust separability, though typologically diverse speech production systems \cite{ladefoged2012vowels} remain to be tested.

\vspace{-2mm}
\subsection{Limitations}
\vspace{-2mm}
Several limitations bound the scope of our conclusions. First, the analysis operates on handcrafted acoustic features; whether learned representations from self-supervised models such as wav2vec 2.0 \cite{baevski2020wav2vec} or HuBERT \cite{hsu2021hubert} preserve the observed dimensional independence remains an open question. Second, the MI estimates characterise \textit{static} feature distributions, and temporal dynamics (e.g., prosodic contours over utterances) may reveal additional cross-dimension dependencies.
Third, the feature groupings are one of several plausible partitions, and partial overlap between groups (e.g., shared formant and voice quality descriptors) means that the measured MI reflects properties of the chosen operationalisation rather than of abstract speech dimensions per se.
Fourth, the corpora are not representative across pathologies or emotional speech paradigms, and confounding factors, recording conditions, native vs.\ non-native speaker status, and corpus-specific mixed effects exert additional influence on the extracted features.
Taken together, these considerations position the present work as demonstrating a methodological framework and providing initial empirical findings that open a new perspective on mutual dependency analysis in speech; individual numerical results should be interpreted accordingly.
Future work will extend this framework to learned neural representations \cite{baevski2020wav2vec,hsu2021hubert} and broader clinical populations \cite{duffy2012motor}.
\vspace{-3mm}


\section{Conclusion}

We presented a bounded information-theoretic framework for quantifying dimensional independence in speech. Across six corpora, emotional, linguistic, and pathological feature sets exhibited, in the considered data, near-zero mutual information (0.10--0.12 nats) with tight estimation bounds ($\Delta < 0.15$ nats), while Source--Filter coupling was substantially higher (0.47 nats). Source--Filter attribution revealed dimension-specific acoustic substrates: emotional features were source-dominated (80\%), whereas linguistic and pathological features showed predominantly filter-based encoding. Validated across six corpora, these findings provide initial empirical support and a principled framework for measuring cross-dimension statistical dependence in speech representations.

\bibliography{mybib}
\bibliographystyle{IEEEtran}

\end{document}